\documentclass[prd,preprint,preprintnumbers,amsmath,amssymb]{revtex4-1}
\usepackage{epsf}
\usepackage{graphicx}% Include figure files
\usepackage{dcolumn}% Align table columns on decimal point
\usepackage{bm}% bold math

\usepackage{centernot}
\newcommand{\nff}{\ensuremath{f}}

\newcommand{\chg}{\ensuremath{e}}
\newcommand{\mgm}{\ensuremath{\mu}}
\newcommand{\elm}{\ensuremath{\epsilon}}
\newcommand{\anm}{\ensuremath{a}}

\newcommand{\bmag}{\ensuremath{\mu_{\text{B}}}}

\def\be{\begin{equation}}
\def\ee{\end{equation}}

\begin{document}
%FTPI-MINN-11/02
%UMN-TH-2935/11
%
\title[]{Electromagnetic properties of massive neutrinos in low-energy elastic neutrino-electron scattering}
\author{Konstantin A. Kouzakov}
\affiliation{Department of Nuclear Physics and Quantum
Theory of Collisions, Faculty of Physics, Lomonosov Moscow State University, Moscow 119991, Russia}%
\email{kouzakov@srd.sinp.msu.ru}
\author{Alexander I. Studenikin}
\affiliation{Department of Theoretical Physics, Faculty of
Physics, Lomonosov Moscow State University, Moscow 119991, Russia}%
\affiliation{Joint Institute for Nuclear Research, Dubna 141980, Moscow Region, Russia}%
\email{studenik@srd.sinp.msu.ru}
\begin{abstract}
A thorough account of electromagnetic interactions of massive neutrinos in the theoretical formulation of low-energy elastic neutrino-electron scattering is given. The formalism of neutrino charge, magnetic, electric, and anapole form factors defined as matrices in the mass basis is employed under the assumption of three-neutrino mixing. The flavor change of neutrinos traveling from the source to the detector is taken into account and the role of the source-detector distance is inspected. The effects of neutrino flavor-transition millicharges and charge radii in the scattering experiments are pointed out.
\end{abstract}

\maketitle

%%
%\begin{keyword}
%%
%%% keywords here, in the form: keyword \sep keyword
%%
%elastic neutrino-electron scattering \sep electromagnetic interactions of neutrino
%%
%%% MSC codes here, in the form: \MSC code \sep code
%%% or \MSC[2008] code \sep code (2000 is the default)
%%
%\end{keyword}
%%


%
\section{Introduction}
\label{intro}
In the standard model neutrinos are massless left-handed fermions which very weakly interact with matter via exchange of the $W^\pm$ and $Z^0$ bosons. The development of our knowledge about neutrino masses and mixing~\cite{Bilenky:2010zza,Xing:2011zza,King:2015review} provides a basis for exploring neutrino properties and interactions beyond the standard model (BSM). In this respect, the study of nonvanishing electromagnetic characteristics of massive neutrinos is of particular interest~\cite{Broggini:2012df,Giunti_RMP2015,giunti16}. It can help not only to shed light on whether neutrinos are Dirac or Majorana particles, but also to constrain the existing BSM theories and/or to hint at new physics.

The possible electromagnetic properties of massive neutrinos include the electric charge (millicharge), the charge radius, the dipole magnetic and electric moments, and the anapole moment. Their effects can be searched in astrophysical environments, where neutrinos propagate in strong magnetic fields and dense matter~\cite{Raffelt:1996wa}, and in laboratory measurements of neutrinos from various sources. In the latter case, a very sensitive and widely used method is provided by the direct measurement of low-energy elastic (anti)neutrino-electron scattering in reactor, accelerator, and solar experiments. A general strategy of such experiments consists in determining deviations of the scattering cross section differential with respect to the energy transfer from the value predicted by the standard model of the electroweak interaction.

So far, neither astrophysical observations nor laboratory measurements have evidenced nonvanishing electromagnetic properties of neutrinos, and only some constraints on their values have been obtained
(the updated list of constraints is given in the review paper \cite{Giunti_RMP2015}). For example, the most stringent constraint on the neutrino millicharge obtained in the scattering experiments is
\begin{equation}
\label{e_nu_reactor_FEA}|e_{{\nu}_e}|\lesssim1.5\times10^{-12}e,
\end{equation}
which has been derived in Ref.~\cite{Studenikin:2013my} from the analysis of the reactor data~\cite{Beda:2012zz} using the free-electron approximation for the differential cross section. If one goes beyond the free-electron approximation and takes into account the binding of electrons to atoms in the detector (the atomic-ionization effect), then one arrives at~\cite{Chen:2014q_nu}
\begin{equation}
\label{e_nu_reactor} |e_{{\nu}_e}|<1.1\times10^{-12}e.
\end{equation}
This bound is orders of magnitude less stringent than those that follow from astrophysics~\cite{Studenikin:2012vi},
$$
|e_{{\nu}_e}|\lesssim 1.3\times10^{-19}e,
$$
and the neutrality of matter~\cite{Raffelt:1999gv},
$$
|e_{{\nu}_e}|\lesssim3\times10^{-21}e.
$$

While neutrinos are generally believed to be electrically neutral particles, they are still expected to have nonzero charge radii. The current constraints from the scattering experiments ($|\langle r_\nu^2\rangle|\lesssim10^{-32}-10^{-31}$\,cm$^2$) differ only by 1 to 2 orders of magnitude from the values calculated within the minimally extended standard model with right-handed neutrinos ($|\langle r_{\nu_\ell}^2\rangle|\sim10^{-33}$\,cm$^2$, $\ell=e,\mu,\tau$)~\cite{Bernabeu:2004jr}. This indicates that the standard model neutrino charge radii could be experimentally tested in the near future.

The experimental bounds for the neutrino millicharges and charge radii discussed above have been obtained under an implicit assumption that neutrinos do not change flavor when scattering on electrons in the detector. However, making this assumption for neutrino-electron scattering due to weak interaction is not necessarily justified in the case of electromagnetic interaction. It means that possible contributions from the neutrino flavor-transition electromagnetic properties should also be taken into account in the data analysis~\footnote{Neutrino-flavor-changing electromagnetic interactions due to neutrino magnetic moments were discussed in Refs.~\cite{Giunti_RMP2015,Fabbricatore:2016nec}}. Therefore, the present work aims at filling the lacuna in the basic theoretical apparatus usually employed for interpretation and analysis of the data of experiments searching for electromagnetic interactions of massive neutrinos in the elastic neutrino-electron scattering.

The paper is organized as follows. Section~\ref{nu_e-m} delivers a brief overview of neutrino electromagnetic form factors. In Sec.~\ref{theory} general formulas for the scattering amplitude and differential cross section are presented. Then, in Sec.~\ref{free-electron}, the free-electron approximation and the stepping formula for the differential cross section are discussed. Section~\ref{L} is devoted to the role of the source-detector distance. The conclusions are drawn in Sec.~\ref{concl}.

\section{Electromagnetic interactions of massive neutrinos}
\label{nu_e-m}
A detailed review of neutrino electromagnetic properties and interactions can be found in Refs.~\cite{Broggini:2012df,Giunti_RMP2015,giunti16}. In this section we briefly outline the general form of the electromagnetic interactions of Dirac and Majorana neutrinos.

There are at least three massive neutrino fields $\nu_{j}$ with respective masses
$m_{j}$ ($j=1,2,3$), which are mixed with the three active flavor neutrinos $\nu_{e}$,
$\nu_{\mu}$, $\nu_{\tau}$. Therefore, the
effective electromagnetic interaction Hamiltonian can be presented as
\begin{equation}
\mathcal{H}_{\rm em}^{(\nu)} = j_{\lambda}^{(\nu)}A^{\lambda}
= \sum_{j,k=1}^{3} \overline{\nu}_{j} \Lambda^{jk}_{\lambda}
\nu_{k} A^{\lambda} , \label{C033}
\end{equation}
where we take into account possible transitions between different
massive neutrinos. The physical effect of
$\mathcal{H}_{\rm em}^{(\nu)}$ is described by the effective
electromagnetic vertex, which in the momentum-space representation depends
only on the four-momentum $q=p_j-p_k$ transferred to the photon
and can be expressed as follows:
\begin{equation}
\Lambda_{\lambda}(q) = \left( \gamma_{\lambda} -
\frac{q_{\lambda}\!\centernot{q}}{q^{2}} \right) \left[ \nff_{Q}(q^{2}) + \nff_{A}(q^{2})
q^{2} \gamma^{5} \right]  - i \sigma_{\lambda\rho} q^{\rho} \left[ \nff_{M}(q^{2}) +
i \nff_{E}(q^{2}) \gamma^{5} \right], \label{C043}
\end{equation}
where $\sigma_{\lambda\rho}=i(\gamma_\lambda\gamma_\rho-\gamma_\rho\gamma_\lambda)/2$.
Here $\Lambda_{\lambda}(q)$ is a $3{\times}3$ matrix in the space
of massive neutrinos expressed in terms of the four Hermitian
$3{\times}3$ matrices of form factors
\begin{equation}
\nff_{Q} = \nff_{Q}^{\dagger}, \qquad \nff_{M} = \nff_{M}^{\dagger},
\qquad \nff_{E} = \nff_{E}^{\dagger}, \qquad \nff_{A} = \nff_{A}^{\dagger},
\label{C044}
\end{equation}
where $Q,M,E,A$ refer, respectively, to the real charge, magnetic, electric, and anapole neutrino form factors. The Lorentz-invariant form of the vertex function~(\ref{C043}) is also consistent with electromagnetic gauge invariance that implies four-current conservation.

For the coupling with a real photon in vacuum ($q^{2}=0$) one has
\begin{equation}
\nff_{Q}^{jk}(0) = \chg_{jk} , \qquad \nff_{M}^{jk}(0) = \mgm_{jk} , \qquad
\nff_{E}^{jk}(0) = \elm_{jk} , \qquad \nff_{A}^{jk}(0) = \anm_{jk} ,
\label{C045}
\end{equation}
where $\chg_{jk}$, $\mgm_{jk}$, $\elm_{jk}$ and $\anm_{jk}$ are,
respectively, the neutrino charge, magnetic moment, electric
moment and anapole moment of diagonal ($j=k$) and transition
($j{\neq}k$) types.

Consider the diagonal case $j=k$. The hermiticity of the electromagnetic
current and the assumption of its invariance under discrete symmetries'
transformations put certain constraints on the form factors, which are in
general different for the Dirac and Majorana neutrinos. In the case of
Dirac neutrinos, the assumption of $CP$ invariance combined with the
hermiticity of the electromagnetic current $J_\mu$ implies that the
electric dipole form factor vanishes, $f_E=0$. At zero momentum transfer
only $f_Q(0)$ and $f_M(0)$---which are called the electric charge and the
magnetic moment, respectively---contribute to the Hamiltonian~(\ref{C033}). The hermiticity also implies that
$f_Q$, $f_A$, and $f_M$ are real. In contrast, in the case of Majorana
neutrinos (regardless of whether $CP$ invariance is violated or not) the
charge, dipole magnetic and electric moments vanish, $f_Q=f_M=f_E=0$, so
that only the anapole moment can be nonvanishing among the
electromagnetic moments. Note that it is possible to
prove~\cite{Nieves:1981zt,Kayser:1982br,Kayser:1984ge} that the existence of a
nonvanishing magnetic moment for a Majorana neutrino would bring about a
clear evidence for $CPT$ violation.

In the off-diagonal case $j\neq k$, the hermiticity by itself does not
imply restrictions on the form factors of Dirac neutrinos. It is possible
to show~\cite{Nieves:1981zt} that, if the assumption of the $CP$ invariance is
added, the form factors $f_Q$, $f_M$, $f_E$, and $f_A$ should have the
same complex phase. For the Majorana neutrino, if $CP$ invariance holds,
there could be either a transition magnetic or a transition electric
moment. Finally, as in the diagonal case, the anapole form factor of a
Majorana neutrino can be nonzero.

%A detailed derivation of the general
%expression of the effective electromagnetic coupling of Dirac and Majorana
%neutrinos in terms of electromagnetic form factors and
%discussion on the properties of the form factors under CP and CPT
%transformations can be found in \cite{Giunti_RMP2015}.

It is usually believed that the
neutrino electric charge $\chg_\nu=\nff_Q(0)$ is zero.
In
the standard model of SU(2)$_L \times $U(1)$_Y$ electroweak
interactions it is possible to get \cite{Foot:1992ui} a general
proof that neutrinos are electrically neutral, which is based on
the requirement of electric charge quantization. The direct
calculations of the neutrino charge in the standard model for
massless (see, for instance,
Refs.~\cite{Bardeen:1972vi,CabralRosetti:1999ad}) and massive neutrinos
\cite{Dvornikov:2003js,Dvornikov:2004sj} also prove that, at
least at the one-loop level, the neutrino electric charge is
gauge independent and vanishes. However, if the neutrino has a
mass, it still may become electrically millicharged. A brief
discussion of different mechanisms for introducing millicharged
particles including neutrinos can be found in
Ref.~\cite{Davidson:2000hf}. In the case of millicharged massive neutrinos, electromagnetic gauge invariance implies that the diagonal electric charges $e_{jj}$ ($j=1,2,3$) are equal~\cite{giunti16}. It should be mentioned that the most
stringent experimental constraints on the electric charge of the
neutrino can be obtained from the neutrality of matter.

Even if the electric charge of a
neutrino is zero, the electric form factor $\nff_Q(q^2)$ can still
contain nontrivial information about neutrino electrostatic properties
\cite{Giunti_RMP2015}. A neutral particle can be characterized by
a superposition of two charge distributions of opposite signs, so
that the particle form factor $\nff_Q(q^2)$ can be nonzero for
$q^2\neq 0$. The mean charge radius (in fact, it is the charged
radius squared) of an electrically neutral neutrino is given by
\begin{equation}\label{nu_cha_rad}
{\langle
r_{\nu}^2\rangle}={6}\left.\frac{d\nff_{Q}(q^2)}{dq^2}\right|_{
q^2=0},
\end{equation}
which is determined by the second term in the power-series
expansion of the neutrino charge form factor.

The most well
studied and understood among the neutrino electromagnetic
characteristics are the dipole magnetic and electric moments,
which are given by the corresponding form factors at $q^2=0$:
\begin{equation}
\mgm_{\nu}=\nff_{M}(0), \qquad \elm_{\nu}=\nff_{E}(0).
\end{equation}
The diagonal magnetic and electric moments of a Dirac neutrino in
the minimally extended standard model with right-handed neutrinos (derived for the first time in Ref.~\cite{Fujikawa:1980yx}) are,
respectively,
\begin{equation}
\label{mu_D}
    \mgm^{D}_{jj}
  = \frac{3e_0 G_F m_{j}}{8\sqrt {2} \pi ^2}\approx 3.2\times 10^{-19}
  \bmag\left(\frac{m_j}{1 \, \text{eV}}\right), \qquad \elm^{D}_{jj}=0,
  \end{equation}
where $\bmag$ is the Bohr magneton. According to Eq.~(\ref{mu_D}) the
value of the neutrino magnetic moment is very small. However, in
many other theoretical frameworks (beyond the minimally extended
standard model) the neutrino magnetic moment can reach values that
are of interest for the next generation of terrestrial experiments
and also accessible for astrophysical observations.
%Note that the
%best laboratory upper limit on a neutrino magnetic moment,
%$\mgm_{\nu} \leq 2.9 \times 10^{-11} \bmag$ (90\% CL), has been obtained by
%the GEMMA collaboration \cite{Beda:2012zz} (see Section~\ref{S004}), and the best
%astrophysical limit is $\mgm_{\nu}\leq 3 \times 10^{-12} \mgm _B$ (90\% CL)
%\cite{Raffelt:1990pj}. The latter bound comes from the constraints on the possible delay of helium ignition of a red giant star in globular clusters due to the cooling induced by the energy loss in the plasmon-decay process $\gamma^*\to\nu\bar{\nu}$ (see Fig.~\ref{E003}).
%%From the lack of observational evidence of this effect,
%%the following limit has been found \cite{Raffelt:1989xu,Raffelt:1990pj,Raffelt:1992pi}
%Recently the limit has been updated in \cite{Viaux:2013hca}
%using state-of-the-art astronomical observations and stellar evolution codes,
%with the results
%%
%\begin{equation}
%\mgm_{\nu}
%<
%\left\{
%%\setlength{\arraycolsep}{2pt}
%\begin{array}{l} \displaystyle
%2.6 \times 10^{-12} \bmag
%\quad
%\text{(68\% CL)}
%,
%\\ \displaystyle
%4.5 \times 10^{-12} \bmag
%\quad
%\text{(95\% CL)}
%.
%\end{array}
%\right.
%\label{E048}
%\end{equation}
%%
%This astrophysical bound on a neutrino
%magnetic moment is applicable to both Dirac and Majorana neutrinos
%and constrains all diagonal and transition
%dipole moments.

The notion of an anapole moment for
a Dirac particle was introduced by Zel'dovich~\cite{Zeldovich:1957zl} after
the discovery of parity violation. In order to understand the
physical characteristics of the anapole moment, it is useful to
consider its effect in the interactions with external
electromagnetic fields. The neutrino anapole moment contributes to
the scattering of neutrinos with charged particles. In order to
discuss its effects, it is convenient to consider strictly neutral
neutrinos with $\nff_{Q}(0)=0$ and define a reduced charge form
factor $\tilde{\nff}_{Q}(q^{2})$ such that
\begin{equation}
\nff_{Q}(q^{2}) = q^2 \, \tilde{\nff}_{Q}(q^{2}) . \label{G069}
\end{equation}
Then, from Eq.~(\ref{nu_cha_rad}), apart from a factor $1/6$, the
reduced charge form factor at $q^2=0$ is just the squared neutrino
charge radius:
\begin{equation}
\tilde{\nff}_{Q}(0) = \frac{1}{6}\,\langle{r}^{2}_\nu\rangle. \label{G070}
\end{equation}
Let us now consider the charge and anapole parts of the neutrino
electromagnetic vertex function, %in Eq.~(\ref{C043}), which can be written
as
\begin{equation}
\Lambda_{\lambda}^{Q,A}(q) = \left( \gamma_{\lambda} q^{2} - q_{\lambda}
\!\centernot{q} \right) \left[ \tilde{\nff}_{Q}(q^{2}) +
\nff_{A}(q^{2}) \gamma^{5} \right] . \label{G071}
\end{equation}
Since for ultrarelativistic neutrinos the effect of $\gamma^{5}$
is only a
sign which depends on the helicity of the neutrino, %(see Eq.~(\ref{K007})),
the phenomenology of neutrino anapole moments is similar to that
of neutrino charge radii.
%Hence, the limits on the neutrino charge
%radii discussed in Section~\ref{S004} apply also to the neutrino
%anapole moments multiplied by a factor of 6.

%
\section{Basic formulas for elastic neutrino-electron scattering}
\label{theory}
We consider the process where an ultrarelativistic neutrino with energy $E_\nu$ originates from a source (reactor, accelerator, the Sun, etc.) and elastically scatters on an electron in a detector at energy-momentum transfer $q=(T,{\bf q})$. If the neutrino is born in the source in the flavor state $|\nu_\ell\rangle$, then its state in the detector is
\begin{equation}
\label{nu_state}|\nu_\ell(L)\rangle=\sum_{k=1}^3U^*_{\ell k}e^{-i\frac{m_k^2}{2E_\nu}L}|\nu_k\rangle,
\end{equation}
where $L$ is the source-detector distance. The matrix element of the transition $\nu_\ell(L)+e^-\to\nu_j+e^-$ due to weak interaction is given by
\begin{eqnarray}
\label{M_weak}\mathcal{M}_j^{(w)}&=&\frac{G_F}{\sqrt{2}}\sum_{k=1}^3U^*_{\ell k}e^{-i\frac{m_k^2}{2E_\nu}L}
\left[(g_V')_{jk}\bar{u}_j\gamma_\lambda(1-\gamma^5)u_kJ_V^\lambda(q)%\right.\nonumber\\%\bar{u}_{e_f}\gamma^\mu u_{e_i}
%&{}&\left.
-(g_A')_{jk}\bar{u}_j\gamma_\lambda(1-\gamma^5)u_kJ_A^\lambda(q)\right],\nonumber\\
%\bar{u}_{e_f}\gamma^\mu\gamma^5 u_{e_i}
\end{eqnarray}
where
$$
(g_V')_{jk}=\delta_{jk}g_V+U^*_{ej}U_{ek}, \qquad (g_A')_{jk}=\delta_{jk}g_A+U^*_{ej}U_{ek},
$$
with $g_V=2\sin^2\theta_W-1/2$, $g_A=-1/2$, and $\bar{u}_j=u_j^\dag\gamma^0$, where $u_j$ ($u_k$) is the bispinor amplitude of the massive neutrino state $|\nu_j\rangle$ ($|\nu_k\rangle$) with four-momentum $p_j$ ($p_k$). The electron transition vector and axial currents in the detector are
\begin{equation}
\label{J_V_A} J_V^\lambda(q)=\langle f|\sum_{d}e^{i{\bf q}\cdot{\bf r}_{d}}\gamma^0_{d}\gamma^\lambda_{d}|i\rangle, \qquad
J_A^\lambda(q)=\langle f|\sum_{d}e^{i{\bf q}\cdot{\bf r}_{d}}\gamma^0_{d}\gamma^\lambda_{d}\gamma^5_{d}|i\rangle,
\end{equation}
where the $d$ sum runs over all electrons in the detector, and $|i\rangle$ and $|f\rangle$ are initial and final states of the detector, such that $\mathcal{E}_f-\mathcal{E}_i=T$, where $\mathcal{E}_i$ and $\mathcal{E}_f$ are the energies of these states.

The matrix element due to electromagnetic interaction is given by
\begin{eqnarray}
\label{M_el-m} \mathcal{M}_j^{(\gamma)}=\mathcal{M}_j^{(Q)}+\mathcal{M}_j^{(\mu)},
\end{eqnarray}
with
\begin{eqnarray}
\label{M_Q}\mathcal{M}_j^{(Q)}&=&\frac{4\pi\alpha}{q^2}\sum_{k=1}^3U^*_{\ell k}e^{-i\frac{m_k^2}{2E_\nu}L}
\bar{u}_j\left(\gamma_\lambda-\frac{q_\lambda\!\centernot q}{q^2}\right)\left[(e_\nu)_{jk}+\frac{q^2}{6}\langle r_\nu^2\rangle_{jk}\right]u_kJ_V^\lambda(q),\\
%\nonumber\\
%\times \bar{u}_{e_f}\gamma^\mu u_{e_i},
%\end{eqnarray}
%%
%%
%\begin{eqnarray}
\label{M_mu}\mathcal{M}_j^{(\mu)}&=&-i\frac{2\pi\alpha}{m_eq^2}\sum_{k=1}^3U^*_{\ell k}e^{-i\frac{m_k^2}{2E_\nu}L}
\bar{u}_j\sigma_{\lambda\rho}q^\rho(\mu_\nu)_{jk}u_kJ_V^\lambda(q),%\bar{u}_{e_f}\gamma^\mu u_{e_i},
\end{eqnarray}
where the neutrino millicharge $e_\nu$ and magnetic moment $\mu_\nu$ are measured in units of $e$ and $\mu_B$, respectively, and the following notation is employed:
$$
(e_\nu)_{jk}=e_{jk}, \qquad \langle r_\nu^2\rangle_{jk}=\langle r^2\rangle_{jk}+6\gamma^5a_{jk}, \qquad  (\mu_\nu)_{jk}=\mu_{jk}+i\gamma^5\epsilon_{jk}.
$$
Taking into account that $\gamma^5|\nu_\ell\rangle=-|\nu_\ell\rangle$, for ultrarelativistic neutrinos we have
$\gamma^5u_k\simeq-u_k$. Therefore, in such a case the effect of $\gamma^5$ in the above formulas is simply a multiplication by a factor of $-1$. Also, in such a case there is no interference between the helicity-conserving ($\mathcal{M}_j^{(w)}$ and $\mathcal{M}_j^{(Q)}$) and helicity-flipping ($\mathcal{M}_j^{(\mu)}$) amplitudes. Combining the helicity-conserving amplitudes, we find
\begin{eqnarray}
\label{M_weak_Q}{\mathcal{M}}_j^{(w,Q)}&=&\mathcal{M}_j^{(w)}+\mathcal{M}_j^{(Q)}\nonumber\\
&=&\frac{G_F}{\sqrt{2}}\sum_{k=1}^3U^*_{\ell k}e^{-i\frac{m_k^2}{2E_\nu}L}
\Big\{\left[(g_V')_{jk}+\tilde{Q}_{jk}\right]\bar{u}_j\gamma_\lambda(1-\gamma^5)u_kJ_V^\lambda(q)\nonumber\\
%\bar{u}_{e_f}\gamma^\mu u_{e_i}
&{}&-(g_A')_{jk}\bar{u}_j\gamma_\lambda(1-\gamma^5)u_kJ_A^\lambda(q)\Big\},
%\bar{u}_{e_f}\gamma^\mu\gamma_5 u_{e_i}
\end{eqnarray}
where
$$
%A_{jk}=U^*_{ej}U_{ek}, \qquad \tilde{A}_{jk}=A_{jk}+B_{jk}, \qquad
\tilde{Q}_{jk}=\frac{2\sqrt{2}\pi\alpha}{G_F}\left[\frac{(e_\nu)_{jk}}{q^2}+\frac{1}{6}\langle r_\nu^2\rangle_{jk}\right].
$$
In Eq.~(\ref{M_weak_Q}), it is taken into account that $q_\lambda J_V^\lambda(q)=0$.

When evaluating the cross section, we neglect the neutrino masses and set $p_j=p'$ and $p_k=p$. Since the final massive state of the neutrino is not resolved in the detector, the differential cross section measured in the scattering experiment is given by
\begin{equation}
\label{cr_sec} \frac{d\sigma}{dT}=\frac{1}{32\pi^2}
\int\limits_{T^2}^{(2E_\nu-T)^2}\frac{d{\bf q}^2}{E_\nu^2}\int\limits_0^{2\pi}d\varphi_{\bf q}\left|\mathcal{M}_{fi}\right|^2\delta(T-\mathcal{E}_f+\mathcal{E}_i),
\end{equation}
%
%%
%\begin{equation}
%\label{cr_sec} \frac{d\sigma}{dT}=\sum_{j=1}^3\frac{d\sigma_j}{dT}.
%\end{equation}
%%
with the following absolute matrix element squared:
\begin{eqnarray}
\label{M_fi} \left|\mathcal{M}_{fi}\right|^2=\sum_{j=1}^3\left\{\left|{\mathcal{M}}_j^{(w,Q)}\right|^2+\left|{\mathcal{M}}_j^{(\mu)}\right|^2\right\},
%&=&\frac{G_F}{\sqrt{2}}\sum_{k=1}^3U^*_{\ell k}e^{-i\frac{m_k^2}{2E_\nu}L}
%\left[(\delta_{jk}g_V+\tilde{A}_{jk})\bar{u}_j\gamma_\mu(1-\gamma^5)u_kJ_V^\mu(q),\right.\nonumber\\%\bar{u}_{e_f}\gamma^\mu u_{e_i}
%&{}&\left.-(\delta_{jk}g_A+A_{jk})\bar{u}_j\gamma_\mu(1-\gamma^5)u_kJ_A^\mu(q)\right],%\bar{u}_{e_f}\gamma^\mu\gamma_5 u_{e_i}
\end{eqnarray}
where, as usual, averaging over initial and summing over final spin polarizations is assumed. The angle $\varphi_{\bf q}$ in Eq.~(\ref{cr_sec}) is the azimuthal angle of the momentum transfer ${\bf q}$ in the spherical coordinate system with the $z$ axis directed along the incident neutrino momentum ${\bf p}$.

Using
\begin{eqnarray}
\frac{1}{4}\,{\rm Sp}\left\{\centernot{\! p}\!'\gamma_{\lambda}(1-\gamma^5)\centernot{\! p}\gamma_{\lambda'}(1-\gamma^5)\right\}
=2[p_{\lambda}p'_{\lambda'}+p'_{\lambda}p_{\lambda'}-(p\cdot p')g_{\lambda\lambda'}%\nonumber\\
-i\varepsilon_{\lambda\rho\lambda'\rho'}p'^{\rho}p^{\rho'}],\nonumber
\end{eqnarray}
where $g_{\lambda\lambda'}$ is the metric tensor and $\varepsilon_{\lambda\rho\lambda'\rho'}$ is the Levi-Civita symbol, we obtain
\begin{eqnarray}
\label{matrx_el_sqrd_SM}
\left|{\mathcal{M}}_{fi}^{(w,Q)}\right|^2&=&\sum_{j=1}^3\left|\tilde{\mathcal{M}}_j^{(w,Q)}\right|^2\nonumber\\
&=& 4G_F^2\Bigg\{C_1\left[2|p\cdot J_V(q)|^2-(p\cdot p')J_V(q)\cdot J_V^*(q)
-i\varepsilon_{\lambda\rho\lambda'\rho'}p'^{\rho}p^{\rho'}J_V^\lambda(q)J_V^{\lambda'*}(q)\right]\nonumber\\
&{}&
+C_2\Big[\left(p\cdot J_A(q)\right)\left(p'\cdot J_A^*(q)\right)+\left(p'\cdot J_A(q)\right)\left(p\cdot J_A^*(q)\right)-(p\cdot p')J_A(q)\cdot J_A^*(q)\nonumber\\
%&{}&\nonumber\\
&{}&-i\varepsilon_{\lambda\rho\lambda'\rho'}p'^{\rho}p^{\rho'}J_A^\lambda(q)J_A^{\lambda'*}(q)\Big]-2{\rm Re}\Big\{C_3\Big[\left(p\cdot J_V(q)\right)\left(p'\cdot J_A^*(q)\right)\nonumber\\
&{}&+\left(p'\cdot J_V(q)\right)\left(p\cdot J_A^*(q)\right)-(p\cdot p')J_V(q)\cdot J_A^*(q)%\nonumber\\&{}&
-i\varepsilon_{\lambda\rho\lambda'\rho'}p'^{\rho}p^{\rho'}J_V^\lambda(q)J_A^{\lambda'*}(q)\Big]\Big\}\Bigg\}.\nonumber\\
\end{eqnarray}
Here
\begin{eqnarray}
C_1&=&\sum_{j,k,k'=1}^3U^*_{\ell k}U_{\ell k'}e^{-i\frac{\delta m_{kk'}^2}{2E_\nu}L}
\left[(g_V')_{jk}+\tilde{Q}_{jk}\right]\left[(g_V')_{jk'}^*+\tilde{Q}_{jk'}^*\right],\label{C1}\\
C_2&=&\sum_{j,k,k'=1}^3U^*_{\ell k}U_{\ell k'}e^{-i\frac{\delta m_{kk'}^2}{2E_\nu}L}
(g_A')_{jk}(g_A')_{jk'}^*,\label{C2}\\
C_3&=&\sum_{j,k,k'=1}^3U^*_{\ell k}U_{\ell k'}e^{-i\frac{\delta m_{kk'}^2}{2E_\nu}L}
\left[(g_V')_{jk}+\tilde{Q}_{jk}\right](g_A')_{jk'}^*,\label{C3}
\end{eqnarray}
with $\delta m_{kk'}^2=m_k^2-m_{k'}^2$.

Using
\begin{eqnarray}
\frac{1}{4}\,{\rm Sp}\left\{\centernot{\! p}\! '\sigma_{\lambda\rho}q^\rho\centernot{\! p}\sigma_{\lambda'\rho'}q^{\rho'}\right\}
=-(p\cdot p')(p_{\lambda}+p'_{\lambda})(p_{\lambda'}+p'_{\lambda'})\nonumber
\end{eqnarray}
and the relations $p+p'=2p-q$, $p\cdot p'=-q^2/2$, and $q_\lambda J_V^\lambda(q)=0$,
we receive
\begin{eqnarray}
\label{matrx_el_sqrd_mu}
\left|{\mathcal{M}}_{fi}^{(\mu)}\right|^2=\sum_{j=1}^3\left|{\mathcal{M}}_j^{(\mu)}\right|^2
= \frac{32\pi^2\alpha^2}{m_e^2|q^2|}|\mu_\nu(L,E_\nu)|^2|p\cdot J_V(q)|^2,
\end{eqnarray}
where the absolute effective magnetic moment squared is given by~\cite{Giunti_RMP2015}
\begin{equation}
\label{eff_mu_nu}
|\mu_\nu(L,E_\nu)|^2=\sum_{j=1}^3\left|\sum_{k=1}^3U^*_{\ell k}e^{-i\frac{m_{k}^2}{2E_\nu}L}(\mu_\nu)_{jk}\right|^2.
\end{equation}

In the case of Dirac antineutrinos, one must make the following substitutions in the above formulas: $U_{\ell k}\to U_{\ell k}^*$, $(g_V')_{jk}\to-(g_V')_{jk}^*$, $(g_A')_{jk}\to-(g_A')_{jk}^*$, $\varepsilon_{\lambda\rho\lambda'\rho'}\to-\varepsilon_{\lambda\rho\lambda'\rho'}$, $(e_{\nu})_{jk}\to(e_{\bar{\nu}})_{jk}=-e_{kj},$ and
$$
\langle r_{\nu}^2\rangle_{jk}\to\langle r_{\bar{\nu}}^2\rangle_{jk}=-\langle r^2\rangle_{kj}+6\gamma^5a_{kj}, \qquad  (\mu_{\nu})_{jk}\to(\mu_{\bar\nu})_{jk}=-\mu_{kj}-i\gamma^5\epsilon_{kj},
$$
where the effect of $\gamma^5$ is a multiplication by a factor of $+1$.

\section{Free-electron approximation}
\label{free-electron}
The simplest model of the electron system in the detector is a free-electron model, where it is assumed that electrons are free and at rest. This approximation is supposed to be applicable if the energy-transfer value $T$ is much larger than the electron binding energy in the detector. The differential cross section~(\ref{cr_sec}) in the case of neutrino scattering on one free electron is
\begin{equation}
\label{cr_sec_FE} \frac{d\sigma}{dT}=\frac{1}{32\pi^2}
\int\limits_{T^2}^{(2E_\nu-T)^2}\frac{d{\bf q}^2}{E_\nu^2}\int\limits_0^{2\pi}d\varphi_{\bf q}\left|\mathcal{M}_{fi}\right|^2\delta(T-\sqrt{{\bf q}^2+m_e^2}+m_e),
\end{equation}
%
%%
%$$
%d\sigma=(2\pi)^4\delta^{(4)}(p'+k'-p-k)|\mathcal{M}_{fi}|^2\frac{1}{4I}\frac{d{\bf p}'}{(2\pi)^32E_\nu'}\frac{d{\bf k}'}{(2\pi)^32E_e'},
%$$
%%
%where $k=(m_e,0)$, $k'=k+q$, $E_\nu'=E_\nu-T$, $E_e'=m_e+T$, $I=E_\nu m_e$.

The free-electron vector and axial currents~(\ref{J_V_A}) are
$$
J_V^\lambda(q)=\frac{1}{2\sqrt{E_e'm_e}}\,\bar{u}_{e}'\gamma^\lambda u_{e}, \qquad
J_A^\lambda(q)=\frac{1}{2\sqrt{E_e'm_e}}\,\bar{u}_{e}'\gamma^\lambda\gamma^5 u_{e},
$$
where $E_e'=m_e+T$ is the final electron energy, and $u_e$ and $u_e'$ are the initial and final electron bispinor amplitudes, which are normalized as $\bar{u}_{e}u_{e}=\bar{u}'_{e}u'_{e}=2m_e$. For the absolute matrix elements squared~(\ref{matrx_el_sqrd_SM}) and~(\ref{matrx_el_sqrd_mu}) one thus has
\begin{eqnarray}
\left|{\mathcal{M}}_{fi}^{(w,Q)}\right|^2&=&\frac{4G_F^2}{E_e'm_e}\,
\Big[(C_1+C_2+2{\rm Re}\left\{C_3\right\})(p\cdot k)(p'\cdot k')\nonumber\\
&{}&+(C_1+C_2-2{\rm Re}\left\{C_3\right\})(p\cdot k')(p'\cdot k)
%\nonumber\\&{}&
+(C_2-C_1)(p\cdot p')m_e^2\Big],\label{M_SM_FE_sqrd}\\
\left|{\mathcal{M}}_{fi}^{(\mu)}\right|^2&=&\frac{32\pi^2\alpha^2}{m_e^3E_e'|q^2|}\,|\mu_\nu(L,E_\nu)|^2(p\cdot k)(p\cdot k'),
\label{M_mu_FE_sqrd}
\end{eqnarray}
where $k=(m_e,0)$ and $k'=k+q$ are the initial and final electron four-momenta.

From conservation of four-momentum, $p+k=p'+k'$, it follows that
$$
p\cdot k=p'\cdot k'=E_\nu m_e, \qquad p\cdot k'=p'\cdot k=(E_\nu-T)m_e, \qquad p\cdot p'=k\cdot k'-m_e^2=Tm_e,
$$
and $q^2=-2m_eT$. Using these relations in Eqs.~(\ref{M_SM_FE_sqrd}) and~(\ref{M_mu_FE_sqrd}), we obtain after performing integrations in Eq.~(\ref{cr_sec_FE}) the differential cross section in the free-electron approximation as
\begin{equation}
\label{cr_sec_FE_1}
\frac{d\sigma^{\rm FE}}{dT}=\frac{d\sigma_{(w,Q)}^{\rm FE}}{dT}+\frac{d\sigma_{(\mu)}^{\rm FE}}{dT},
\end{equation}
with
\begin{eqnarray}
\frac{d\sigma_{(w,Q)}^{\rm FE}}{dT}&=&\frac{G_F^2m_e}{2\pi}\left[C_1+C_2+2{\rm Re}\left\{C_3\right\}+(C_1+C_2-2{\rm Re}\left\{C_3\right\})\left(1-\frac{T}{E_\nu}\right)^2\right.
%\nonumber\\&{}&
\nonumber\\
&{}&\left.
+(C_2-C_1)\frac{Tm_e}{E_\nu^2}\right],\\
\frac{d\sigma_{(\mu)}^{\rm FE}}{dT}&=&\frac{\pi\alpha^2}{m_e^2}\,|\mu_\nu(L,E_\nu)|^2\left(\frac{1}{T}-\frac{1}{E_\nu}\right).
\end{eqnarray}

When the energy-transfer value $T$ is comparable to the electron binding energy, the free-electron approximation becomes not generally valid anymore. In particular, for atomic electrons it was found that as the value of $T$ decreases the contribution to the cross section associated with the neutrino millicharge exhibits strong enhancement as compared to the free-electron case~\cite{Chen:2014q_nu}. This is the so-called atomic ionization effect, which is observed for ultrarelativistic charged projectiles and which can be estimated within the equivalent photon approximation. At the same time, if the neutrino millicharges are zero, i.e., $e_{jk}=0$, the cross section for neutrino scattering on atomic electrons is well approximated by the stepping formula
\begin{equation}
\label{cr_sec_step}
\frac{d\sigma}{dT}=\frac{d\sigma^{\rm FE}}{dT}\sum_{\beta}n_\beta\theta(T-\varepsilon_\beta),
\end{equation}
where $n_\beta$ and $\varepsilon_\beta$ are the number and binding energy of electrons in the (sub)shell $\beta$. The stepping approximation was first introduced in Ref.~\cite{Kopeikin:1997step} on the basis of numerical calculations for the case of an iodine atomic target, and later it was supported by a general theoretical analysis~\cite{Kouzakov:2011vx,Kouzakov:2014lka}. Notable deviations of the weak and magnetic cross sections from the stepping formula~(\ref{cr_sec_step}) are found only close to the ionization threshold~\cite{Chen:2013lba,Kouzakov:2014pepan_lett}, where the cross-section values decrease relative to the free-electron approximation. The latter behavior is attributed to the effects of electron-electron correlations in atoms~\cite{Kouzakov:2014lka}.

\section{The role of neutrino flavor oscillations}
\label{L}
It is clear that the manifestation of the neutrino electromagnetic properties in the discussed scattering process depends on the neutrino state $\nu_\ell(L)$ in the detector. Neutrino flavor oscillations are determined by the source-detector distance and the neutrino energy. Below we inspect their impact on the general formulas presented in Sec.~\ref{theory}.

Introducing the flavor transition amplitude and probability,
$$
\mathcal{A}_{\nu_\ell\to\nu_{\ell'}}(L,E_\nu)=\langle\nu_{\ell'}|\nu_\ell(L)\rangle=\sum_{k=1}^3U^*_{\ell k}U_{\ell' k}e^{-i\frac{m_{k}^2}{2E_\nu}L}, \qquad P_{\nu_\ell\to\nu_{\ell'}}(L,E_\nu)=|\mathcal{A}_{\nu_\ell\to\nu_{\ell'}}(L,E_\nu)|^2,
%P_{\nu_\ell\to\nu_e}(L,E_\nu)=\sum_{k,k'=1}^3U^*_{\ell k}U_{\ell k'}U^*_{e k'}U_{ek}e^{-i\frac{\delta m_{kk'}^2}{2E_\nu}L},
%\qquad \tilde{Q}^{(\ell)}_j=\sum_{k=1}^3U^*_{\ell k}e^{-i\frac{m_{k}^2}{2E_\nu}L}\tilde{Q}_{jk},
$$
we arrive at
\begin{eqnarray}
C_1&=&g_V^2+2g_VP_{\nu_\ell\to\nu_e}(L,E_\nu)+P_{\nu_\ell\to\nu_e}(L,E_\nu)+
2g_V\sum_{\ell',\ell''=e,\mu,\tau}\mathcal{A}_{\nu_\ell\to\nu_{\ell'}}(L,E_\nu)\mathcal{A}_{\nu_\ell\to\nu_{\ell''}}^*(L,E_\nu)\tilde{Q}_{\ell''\ell'}
\nonumber\\&{}&
+2{\rm Re}\left\{\mathcal{A}_{\nu_\ell\to\nu_{e}}^*(L,E_\nu)\sum_{\ell'=e,\mu,\tau}\mathcal{A}_{\nu_\ell\to\nu_{\ell'}}(L,E_\nu)\tilde{Q}_{e\ell'}\right\}
\nonumber\\
&{}&
+\sum_{\ell',\ell'',\ell'''=e,\mu,\tau}\mathcal{A}_{\nu_\ell\to\nu_{\ell'}}(L,E_\nu)\mathcal{A}_{\nu_\ell\to\nu_{\ell''}}^*(L,E_\nu)\tilde{Q}_{\ell''\ell'''}\tilde{Q}_{\ell'''\ell'},\label{C1_flavor}\\
C_2&=&g_A^2+2g_AP_{\nu_\ell\to\nu_e}(L,E_\nu)+P_{\nu_\ell\to\nu_e}(L,E_\nu),\label{C2_flavor}\\
C_3&=&g_Vg_A+(g_V+g_A+1)P_{\nu_\ell\to\nu_e}(L,E_\nu)+
g_A\sum_{\ell',\ell''=e,\mu,\tau}\mathcal{A}_{\nu_\ell\to\nu_{\ell'}}(L,E_\nu)\mathcal{A}_{\nu_\ell\to\nu_{\ell''}}^*(L,E_\nu)\tilde{Q}_{\ell''\ell'}\nonumber\\
&{}&
+\mathcal{A}_{\nu_\ell\to\nu_{e}}^*(L,E_\nu)\sum_{\ell'=e,\mu,\tau}\mathcal{A}_{\nu_\ell\to\nu_{\ell'}}(L,E_\nu)\tilde{Q}_{e\ell'},\label{C3_flavor}
\end{eqnarray}
with
$$
\tilde{Q}_{\ell'\ell}=\sum_{j,k=1}^3 U_{\ell' j}U_{\ell k}^*\tilde{Q}_{jk}=\frac{2\sqrt{2}\pi\alpha}{G_F}\left[\frac{(e_\nu)_{\ell'\ell}}{q^2}
+\frac{1}{6}\langle r^2_\nu\rangle_{\ell'\ell}\right],
$$
where
$$
(e_\nu)_{\ell'\ell}=\sum_{j,k=1}^3 U_{\ell' j}U_{\ell k}^*(e_\nu)_{jk} \quad \text{and} \quad
\langle r^2_\nu\rangle_{\ell'\ell}=\sum_{j,k=1}^3 U_{\ell' j}U_{\ell k}^*\langle r^2_\nu\rangle_{jk}
$$
are the neutrino millicharge and charge radius in the flavor basis. In Eq.~(\ref{C1_flavor}), it is taken into account that
$\tilde{Q}_{\ell\ell'}=\tilde{Q}_{\ell'\ell}^*$ due to hermiticity of the neutrino electromagnetic form factors $f_Q$ and $f_A$.

Let us consider two typical cases of the scattering experiments: (i) short-baseline (reactor and accelerator neutrino experiments) and (ii) long-baseline (solar neutrino experiments). In the short-baseline experiments the effect of neutrino flavor change is insignificant, so that to a close approximation the neutrino flavor in the detector is the same as in the source. On the contrary, in the long-baseline experiments neutrinos can change their flavor many times when propagating from the source to the detector. Due to the finite energy resolution of the detector the interference effects in neutrino flavor oscillations over long distances appear to be washed out. In what follows, we formulate these behaviors mathematically.

In the short-baseline case we have $L\ll L_{kk'}={2E_\nu}/|\delta m_{kk'}^2|$ for any $k$ and $k'$. This validates the approximation
$e^{-i(\delta m^2_{kk'}/2E_\nu)L}=1$. Using it, we find
$$
\mathcal{A}_{\nu_\ell\to\nu_{\ell'}}(L,E_\nu)\mathcal{A}_{\nu_\ell\to\nu_{\ell''}}^*(L,E_\nu)=\delta_{\ell\ell'}\delta_{\ell\ell''}, \qquad P_{\nu_\ell\to\nu_e}(L,E_\nu)=\delta_{\ell e}.
$$
Therefore, from Eqs.~(\ref{C1_flavor}),~(\ref{C2_flavor}), and~(\ref{C3_flavor}) we derive, respectively,
\begin{eqnarray}
\label{C1_flavor_short_baseline}
C_1&=&(g_V+\delta_{\ell e}+\tilde{Q}_{\ell\ell})^2+\sum_{\ell'=e,\mu,\tau}(1-\delta_{\ell'\ell})\left|\tilde{Q}_{\ell'\ell}\right|^2,\\
C_2&=&(g_A+\delta_{\ell e})^2,\\
C_3&=&(g_V+\delta_{\ell e})(g_A+\delta_{\ell e})+(g_A+\delta_{\ell e})\tilde{Q}_{\ell\ell}.
\end{eqnarray}
This shows that the weak-electromagnetic interference term contains only flavor-diagonal neutrino millicharges and charge radii.

For the absolute effective magnetic moment squared~(\ref{eff_mu_nu_short-baseline}) we get
\begin{equation}
\label{eff_mu_nu_short-baseline}
|\mu_\nu(L,E_\nu)|^2=\sum_{j=1}^3\sum_{k,k'=1}^3U^*_{\ell k}U_{\ell k'}(\mu_\nu)_{jk}(\mu_\nu)_{jk'}^*=\sum_{\ell'=e,\mu,\tau}
\left|(\mu_\nu)_{\ell'\ell}\right|^2,
\end{equation}
where
$$
(\mu_\nu)_{\ell'\ell}=\sum_{j,k=1}^3U^*_{\ell k}U_{\ell' j}(\mu_\nu)_{jk}
$$
is the effective magnetic moment in the flavor basis.

In the long-baseline case we have $L\gg L_{kk'}={2E_\nu}/|\delta m_{kk'}^2|$ for any $k$ and $k'$. Taking into account the decoherence effects, we can set $e^{-i(\delta m^2_{kk'}/2E_\nu)L}=\delta_{kk'}$ in Eqs.~(\ref{C1}),~(\ref{C2}), and~(\ref{C3}). Hence,
we get
\begin{eqnarray}
C_1&=&g_V^2+2g_VP_{\nu_\ell\to\nu_e}+P_{\nu_\ell\to\nu_e}
+\sum_{j,k=1}^3|U_{\ell k}|^2\left|\tilde{Q}_{jk}\right|^2%\nonumber\\&{}&
+
2g_V\sum_{j=1}^3|U_{\ell j}|^2\tilde{Q}_{jj}\nonumber\\&{}&
+2\sum_{j,k=1}^3|U_{\ell k}|^2{\rm Re}\left\{U_{ej}U^*_{ek}\tilde{Q}_{jk}\right\},\label{C1_long-baseline}\\
C_2&=&g_A^2+2g_AP_{\nu_\ell\to\nu_e}+P_{\nu_\ell\to\nu_e},\label{C2_long-baseline}\\
C_3&=&g_Vg_A+(g_V+g_A+1)P_{\nu_\ell\to\nu_e}+
g_A\sum_{j=1}^3|U_{\ell j}|^2\tilde{Q}_{jj}%\nonumber\\&{}&
+2\sum_{j,k=1}^3|U_{\ell k}|^2 U_{ej}U^*_{ek}\tilde{Q}_{jk},\label{C3_long-baseline}
\end{eqnarray}
where the flavor transition probability
$$
P_{\nu_\ell\to\nu_e}=\sum_{k=1}^3|U_{\ell k}|^2|U_{ek}|^2
$$
does not depend both on the source-detector distance and on the neutrino energy.

For the absolute effective magnetic moment squared~(\ref{eff_mu_nu_short-baseline}) we find
\begin{equation}
\label{eff_mu_nu_long-baseline}
|\mu_\nu(L,E_\nu)|^2=\sum_{j,k=1}^3\left|U_{\ell k}\right|^2\left|(\mu_\nu)_{jk}\right|^2.
\end{equation}
As in the case of Eq.~(\ref{eff_mu_nu_short-baseline}), it is independent of the source-detector distance and neutrino energy.

\section{Summary and concluding remarks}
\label{concl}
We have considered theoretically the low-energy elastic neutrino-electron scattering, taking into account electromagnetic interactions of massive neutrinos. General formulas for the calculation of differential cross sections have been derived in the framework of three-neutrino mixing. The free-electron approximation and stepping formula for the differential cross sections have been discussed. The role of neutrino flavor oscillations has been outlined depending on the source-detector distance.

In contrast to the previous works on neutrino electromagnetic interactions in the processes of elastic neutrino-electron scattering, in the present study the cross section is formulated not in terms of some effective electromagnetic characteristics of the neutrino state $\nu_\ell(L)$ in a detector, but in terms of $3\times3$ matrices of neutrino electromagnetic form factors. It was shown that in the short-baseline experiments one studies these form factors in the flavor basis rather than in the fundamental, mass basis, which is more convenient for interpreting the results of the long-baseline experiments. %This has the following important consequences.

So far, in the analysis of the data of experiments on elastic neutrino-electron scattering the effect of the neutrino charge radius has been considered to be only a shift of the vector coupling constant, $g_V\to g_V+\frac{2}{3}\,M_W^2\langle r_{\nu_\ell(L)}^2\rangle\sin^2\theta_W$ (see, for instance, Ref.~\cite{Vogel:1989iv}). However, one thus misses certain contributions to the cross section from the neutrino charge radius matrix, namely those which do not interfere with the weak-interaction contribution. For example, the current most stringent constraints on the charge radius of the electron antineutrino obtained in this way are
%$|\langle r_{{\nu}_\mu}^2\rangle|<1.2\times10^{-32}$\,cm$^2$.
\begin{equation}
\label{r_nu_TEXONO}
-4.2\times10^{-32}~{\rm cm}^2<\langle r_{{\bar\nu}_e}^2\rangle<6.6\times10^{-32}~{\rm cm}^2,
\end{equation}
which are due to the TEXONO experiment with reactor antineutrinos~\cite{Deniz:2009mu}. The leading role in the derivation of the above bounds is played by the interference term $\propto g_V\langle r_{{\bar\nu}_e}^2\rangle$ in the cross section, while the term $\propto|\langle r_{{\bar\nu}_e}^2\rangle|^2$ is subsidiary. At the same time, according to Eq.~(\ref{C1_flavor_short_baseline}), there is also the term $\propto|\langle r_{{\bar\nu}_e\to{\bar\nu}_\mu}^2\rangle|^2+|\langle r_{{\bar\nu}_e\to{\bar\nu}_\tau}^2\rangle|^2$, where $\langle r_{{\bar\nu}_e\to{\bar\nu}_\mu}^2\rangle=\langle r_{\bar\nu}^2\rangle_{\mu e}$ and $\langle r_{{\bar\nu}_e\to{\bar\nu}_\tau}^2\rangle=\langle r_{\bar\nu}^2\rangle_{\tau e}$ are the transition charge radii in the flavor basis. The contributions from the flavor-transition charge radii do not interfere with the contribution from weak interaction. Hence, these charge radii can have values $\sim10^{-32}~{\rm cm}^2$, without notably affecting the constraints~(\ref{r_nu_TEXONO}).

Finally, some comments should be made regarding contributions to the cross section from neutrino millicharges.
%The most stringent constraint on the $e_\nu$ value from the scattering experiments has been obtained in Ref.~\cite{Chen:2014q_nu} from the analysis of the reactor data~\cite{Beda:2012zz}:
%%
%\begin{equation}
%\label{e_nu_reactor} |e_{\bar{\nu}_e}|<1.1\times10^{-12}e.
%\end{equation}
%%
The bound~(\ref{e_nu_reactor}) has been derived in the region of small $T$ values, where the weak-millicharge interference term is not important and where the atomic-ionization effect is to be taken into account. It follows from Eq.~(\ref{C1_flavor_short_baseline}) that one must understand $|e_{{\nu}_e}|$ in Eq.~(\ref{e_nu_reactor}) as
$$
|e_{{\nu}_e}|=\sqrt{|(e_{{\nu}})_{ee}|^2+|(e_{{\nu}})_{\mu e}|^2+|(e_{{\nu}})_{\tau e}|^2}.
$$
In other words, the flavor-transition millicharges $(e_{{\nu}})_{\mu e}$ and $(e_{{\nu}})_{\tau e}$ also contribute to the cross section in addition to the usual, flavor-diagonal millicharge $(e_{{\nu}})_{ee}$.

%
%\section*{Acknowledgements}
%
%We thank Victor B. Brudanin and Alexander S. Starostin for useful
%discussions. We are grateful to Mikhail B. Voloshin for valuable
%comments.
\begin{acknowledgments}
We thank Carlo Giunti, Anatoly Borisov, Alexey Lokhov, and Dmitry Medvedev for useful discussions.
This work was supported by the Russian Foundation for Basic Research under grants
No.~16-02-01023\,A and No.~17-52-53133\,GFEN\_a.
\end{acknowledgments}
\bibliography{emp_prd}

\begin{thebibliography}{33}%
\makeatletter
\providecommand \@ifxundefined [1]{%
 \@ifx{#1\undefined}
}%
\providecommand \@ifnum [1]{%
 \ifnum #1\expandafter \@firstoftwo
 \else \expandafter \@secondoftwo
 \fi
}%
\providecommand \@ifx [1]{%
 \ifx #1\expandafter \@firstoftwo
 \else \expandafter \@secondoftwo
 \fi
}%
\providecommand \natexlab [1]{#1}%
\providecommand \enquote  [1]{``#1''}%
\providecommand \bibnamefont  [1]{#1}%
\providecommand \bibfnamefont [1]{#1}%
\providecommand \citenamefont [1]{#1}%
\providecommand \href@noop [0]{\@secondoftwo}%
\providecommand \href [0]{\begingroup \@sanitize@url \@href}%
\providecommand \@href[1]{\@@startlink{#1}\@@href}%
\providecommand \@@href[1]{\endgroup#1\@@endlink}%
\providecommand \@sanitize@url [0]{\catcode `\\12\catcode `\$12\catcode
  `\&12\catcode `\#12\catcode `\^12\catcode `\_12\catcode `\%12\relax}%
\providecommand \@@startlink[1]{}%
\providecommand \@@endlink[0]{}%
\providecommand \url  [0]{\begingroup\@sanitize@url \@url }%
\providecommand \@url [1]{\endgroup\@href {#1}{\urlprefix }}%
\providecommand \urlprefix  [0]{URL }%
\providecommand \Eprint [0]{\href }%
\providecommand \doibase [0]{http://dx.doi.org/}%
\providecommand \selectlanguage [0]{\@gobble}%
\providecommand \bibinfo  [0]{\@secondoftwo}%
\providecommand \bibfield  [0]{\@secondoftwo}%
\providecommand \translation [1]{[#1]}%
\providecommand \BibitemOpen [0]{}%
\providecommand \bibitemStop [0]{}%
\providecommand \bibitemNoStop [0]{.\EOS\space}%
\providecommand \EOS [0]{\spacefactor3000\relax}%
\providecommand \BibitemShut  [1]{\csname bibitem#1\endcsname}%
\let\auto@bib@innerbib\@empty
%</preamble>
\bibitem [{\citenamefont {Bilenky}(2010)}]{Bilenky:2010zza}%
  \BibitemOpen
  \bibfield  {author} {\bibinfo {author} {\bibfnamefont {S.}~\bibnamefont
  {Bilenky}},\ }\href {\doibase 10.1007/978-3-642-14043-3} {\emph {\bibinfo
  {title} {{Introduction to the Physics of Massive and Mixed Neutrinos}}}}\
  (\bibinfo  {publisher} {Springer, New York},\ \bibinfo {year}
  {2010})\BibitemShut {NoStop}%
\bibitem [{\citenamefont {Xing}\ and\ \citenamefont
  {Zhou}(2011)}]{Xing:2011zza}%
  \BibitemOpen
  \bibfield  {author} {\bibinfo {author} {\bibfnamefont {Z.-z.}\ \bibnamefont
  {Xing}}\ and\ \bibinfo {author} {\bibfnamefont {S.}~\bibnamefont {Zhou}},\
  }\href@noop {} {\emph {\bibinfo {title} {{Neutrinos in Particle Physics,
  Astronomy and Cosmology}}}}\ (\bibinfo  {publisher} {Zhejiang University
  Press, Zhejiang},\ \bibinfo {year} {2011})\BibitemShut {NoStop}%
\bibitem [{\citenamefont {King}(2015)}]{King:2015review}%
  \BibitemOpen
  \bibfield  {author} {\bibinfo {author} {\bibfnamefont {S.~F.}\ \bibnamefont
  {King}},\ }\href@noop {} {\bibfield  {journal} {\bibinfo  {journal} {J. Phys.
  G}\ }\textbf {\bibinfo {volume} {42}},\ \bibinfo {pages} {123001} (\bibinfo
  {year} {2015})},\ \Eprint {http://arxiv.org/abs/arXiv:1510.02091}
  {arXiv:1510.02091 [hep-ph]} \BibitemShut {NoStop}%
\bibitem [{\citenamefont {Broggini}\ \emph {et~al.}(2012)\citenamefont
  {Broggini}, \citenamefont {Giunti},\ and\ \citenamefont
  {Studenikin}}]{Broggini:2012df}%
  \BibitemOpen
  \bibfield  {author} {\bibinfo {author} {\bibfnamefont {C.}~\bibnamefont
  {Broggini}}, \bibinfo {author} {\bibfnamefont {C.}~\bibnamefont {Giunti}}, \
  and\ \bibinfo {author} {\bibfnamefont {A.}~\bibnamefont {Studenikin}},\
  }\href@noop {} {\bibfield  {journal} {\bibinfo  {journal} {Adv. High Energy
  Phys.}\ }\textbf {\bibinfo {volume} {2012}},\ \bibinfo {pages} {459526}
  (\bibinfo {year} {2012})},\ \Eprint {http://arxiv.org/abs/arXiv:1207.3980}
  {arXiv:1207.3980 [hep-ph]} \BibitemShut {NoStop}%
\bibitem [{\citenamefont {Giunti}\ and\ \citenamefont
  {Studenikin}(2015)}]{Giunti_RMP2015}%
  \BibitemOpen
  \bibfield  {author} {\bibinfo {author} {\bibfnamefont {C.}~\bibnamefont
  {Giunti}}\ and\ \bibinfo {author} {\bibfnamefont {A.}~\bibnamefont
  {Studenikin}},\ }\href@noop {} {\bibfield  {journal} {\bibinfo  {journal}
  {Rev. Mod. Phys.}\ }\textbf {\bibinfo {volume} {87}},\ \bibinfo {pages} {531}
  (\bibinfo {year} {2015})},\ \Eprint {http://arxiv.org/abs/arXiv:1403.6344}
  {arXiv:1403.6344 [hep-ph]} \BibitemShut {NoStop}%
\bibitem [{\citenamefont {Giunti}\ \emph {et~al.}(2016)\citenamefont {Giunti},
  \citenamefont {Kouzakov}, \citenamefont {Li}, \citenamefont {Lokhov},
  \citenamefont {Studenikin},\ and\ \citenamefont {Zhou}}]{giunti16}%
  \BibitemOpen
  \bibfield  {author} {\bibinfo {author} {\bibfnamefont {C.}~\bibnamefont
  {Giunti}}, \bibinfo {author} {\bibfnamefont {K.~A.}\ \bibnamefont
  {Kouzakov}}, \bibinfo {author} {\bibfnamefont {Y.-F.}\ \bibnamefont {Li}},
  \bibinfo {author} {\bibfnamefont {A.~V.}\ \bibnamefont {Lokhov}}, \bibinfo
  {author} {\bibfnamefont {A.~I.}\ \bibnamefont {Studenikin}}, \ and\ \bibinfo
  {author} {\bibfnamefont {S.}~\bibnamefont {Zhou}},\ }\href@noop {} {\bibfield
   {journal} {\bibinfo  {journal} {Ann. Phys. (Berlin)}\ }\textbf {\bibinfo
  {volume} {528}},\ \bibinfo {pages} {198} (\bibinfo {year} {2016})},\ \Eprint
  {http://arxiv.org/abs/arXiv:1506.05387} {arXiv:1506.05387 [hep-ph]}
  \BibitemShut {NoStop}%
\bibitem [{\citenamefont {Raffelt}(1996)}]{Raffelt:1996wa}%
  \BibitemOpen
  \bibfield  {author} {\bibinfo {author} {\bibfnamefont {G.}~\bibnamefont
  {Raffelt}},\ }\href@noop {} {\emph {\bibinfo {title} {{Stars as Laboratories
  for Fundamental Physics: The Astrophysics of Neutrinos, Axions, and Other
  Weakly Interacting Particles}}}}\ (\bibinfo  {publisher} {University of
  Chicago Press, Chicago},\ \bibinfo {year} {1996})\BibitemShut {NoStop}%
\bibitem [{\citenamefont {Studenikin}(2014)}]{Studenikin:2013my}%
  \BibitemOpen
  \bibfield  {author} {\bibinfo {author} {\bibfnamefont {A.}~\bibnamefont
  {Studenikin}},\ }\href@noop {} {\bibfield  {journal} {\bibinfo  {journal}
  {Europhys. Lett.}\ }\textbf {\bibinfo {volume} {107}},\ \bibinfo {pages}
  {21001} (\bibinfo {year} {2014})},\ \Eprint
  {http://arxiv.org/abs/arXiv:1302.1168} {arXiv:1302.1168 [hep-ph]}
  \BibitemShut {NoStop}%
\bibitem [{\citenamefont {Beda}\ \emph {et~al.}(2012)\citenamefont {Beda} \emph
  {et~al.}}]{Beda:2012zz}%
  \BibitemOpen
  \bibfield  {author} {\bibinfo {author} {\bibfnamefont {A.~G.}\ \bibnamefont
  {Beda}} \emph {et~al.} (\bibinfo {collaboration} {GEMMA Collaboration}),\
  }\href {\doibase 10.1155/2012/350150} {\bibfield  {journal} {\bibinfo
  {journal} {Adv. High Energy Phys.}\ }\textbf {\bibinfo {volume} {2012}},\
  \bibinfo {pages} {350150} (\bibinfo {year} {2012})}\BibitemShut {NoStop}%
\bibitem [{\citenamefont {Chen}\ \emph
  {et~al.}(2014{\natexlab{a}})\citenamefont {Chen}, \citenamefont {Chi},
  \citenamefont {Li}, \citenamefont {Liu}, \citenamefont {Singh}, \citenamefont
  {Wong}, \citenamefont {Wu},\ and\ \citenamefont {Wu}}]{Chen:2014q_nu}%
  \BibitemOpen
  \bibfield  {author} {\bibinfo {author} {\bibfnamefont {J.-W.}\ \bibnamefont
  {Chen}}, \bibinfo {author} {\bibfnamefont {H.-C.}\ \bibnamefont {Chi}},
  \bibinfo {author} {\bibfnamefont {H.-B.}\ \bibnamefont {Li}}, \bibinfo
  {author} {\bibfnamefont {C.-P.}\ \bibnamefont {Liu}}, \bibinfo {author}
  {\bibfnamefont {L.}~\bibnamefont {Singh}}, \bibinfo {author} {\bibfnamefont
  {H.~T.}\ \bibnamefont {Wong}}, \bibinfo {author} {\bibfnamefont {C.-L.}\
  \bibnamefont {Wu}}, \ and\ \bibinfo {author} {\bibfnamefont {C.-P.}\
  \bibnamefont {Wu}},\ }\href {\doibase 10.1103/PhysRevD.90.011301} {\bibfield
  {journal} {\bibinfo  {journal} {Phys. Rev. D}\ }\textbf {\bibinfo {volume}
  {90}},\ \bibinfo {pages} {011301} (\bibinfo {year} {2014}{\natexlab{a}})},\
  \Eprint {http://arxiv.org/abs/arXiv:1405.7168 [hep-ph]} {arXiv:1405.7168
  [hep-ph]} \BibitemShut {NoStop}%
\bibitem [{\citenamefont {Studenikin}\ and\ \citenamefont
  {Tokarev}(2014)}]{Studenikin:2012vi}%
  \BibitemOpen
  \bibfield  {author} {\bibinfo {author} {\bibfnamefont {A.}~\bibnamefont
  {Studenikin}}\ and\ \bibinfo {author} {\bibfnamefont {I.}~\bibnamefont
  {Tokarev}},\ }\href@noop {} {\bibfield  {journal} {\bibinfo  {journal} {Nucl.
  Phys.}\ }\textbf {\bibinfo {volume} {B884}},\ \bibinfo {pages} {396}
  (\bibinfo {year} {2014})},\ \Eprint {http://arxiv.org/abs/arXiv:1209.3245}
  {arXiv:1209.3245 [hep-ph]} \BibitemShut {NoStop}%
\bibitem [{\citenamefont {Raffelt}(1999)}]{Raffelt:1999gv}%
  \BibitemOpen
  \bibfield  {author} {\bibinfo {author} {\bibfnamefont {G.~G.}\ \bibnamefont
  {Raffelt}},\ }\href {\doibase 10.1016/S0370-1573(99)00074-5} {\bibfield
  {journal} {\bibinfo  {journal} {Phys. Rep.}\ }\textbf {\bibinfo {volume}
  {320}},\ \bibinfo {pages} {319} (\bibinfo {year} {1999})}\BibitemShut
  {NoStop}%
\bibitem [{\citenamefont {Binosi}\ \emph {et~al.}(2005)\citenamefont {Binosi},
  \citenamefont {Bernabeu},\ and\ \citenamefont
  {Papavassiliou}}]{Bernabeu:2004jr}%
  \BibitemOpen
  \bibfield  {author} {\bibinfo {author} {\bibfnamefont {D.}~\bibnamefont
  {Binosi}}, \bibinfo {author} {\bibfnamefont {J.}~\bibnamefont {Bernabeu}}, \
  and\ \bibinfo {author} {\bibfnamefont {J.}~\bibnamefont {Papavassiliou}},\
  }\href@noop {} {\bibfield  {journal} {\bibinfo  {journal} {Nucl. Phys.}\
  }\textbf {\bibinfo {volume} {B716}},\ \bibinfo {pages} {352} (\bibinfo {year}
  {2005})},\ \Eprint {http://arxiv.org/abs/arXiv:hep-ph/0405288}
  {arXiv:hep-ph/0405288} \BibitemShut {NoStop}%
\bibitem [{Note1()}]{Note1}%
  \BibitemOpen
  \bibinfo {note} {Neutrino-flavor-changing electromagnetic interactions due to
  neutrino magnetic moments were discussed in Refs.~\cite
  {Giunti_RMP2015,Fabbricatore:2016nec}}\BibitemShut {NoStop}%
\bibitem [{\citenamefont {Nieves}(1982)}]{Nieves:1981zt}%
  \BibitemOpen
  \bibfield  {author} {\bibinfo {author} {\bibfnamefont {J.~F.}\ \bibnamefont
  {Nieves}},\ }\href@noop {} {\bibfield  {journal} {\bibinfo  {journal} {Phys.
  Rev. D}\ }\textbf {\bibinfo {volume} {26}},\ \bibinfo {pages} {3152}
  (\bibinfo {year} {1982})}\BibitemShut {NoStop}%
\bibitem [{\citenamefont {Kayser}(1982)}]{Kayser:1982br}%
  \BibitemOpen
  \bibfield  {author} {\bibinfo {author} {\bibfnamefont {B.}~\bibnamefont
  {Kayser}},\ }\href@noop {} {\bibfield  {journal} {\bibinfo  {journal} {Phys.
  Rev. D}\ }\textbf {\bibinfo {volume} {26}},\ \bibinfo {pages} {1662}
  (\bibinfo {year} {1982})}\BibitemShut {NoStop}%
\bibitem [{\citenamefont {Kayser}(1984)}]{Kayser:1984ge}%
  \BibitemOpen
  \bibfield  {author} {\bibinfo {author} {\bibfnamefont {B.}~\bibnamefont
  {Kayser}},\ }\href@noop {} {\bibfield  {journal} {\bibinfo  {journal} {Phys.
  Rev. D}\ }\textbf {\bibinfo {volume} {30}},\ \bibinfo {pages} {1023}
  (\bibinfo {year} {1984})}\BibitemShut {NoStop}%
\bibitem [{\citenamefont {Foot}\ \emph {et~al.}(1993)\citenamefont {Foot},
  \citenamefont {Lew},\ and\ \citenamefont {Volkas}}]{Foot:1992ui}%
  \BibitemOpen
  \bibfield  {author} {\bibinfo {author} {\bibfnamefont {R.}~\bibnamefont
  {Foot}}, \bibinfo {author} {\bibfnamefont {H.}~\bibnamefont {Lew}}, \ and\
  \bibinfo {author} {\bibfnamefont {R.~R.}\ \bibnamefont {Volkas}},\
  }\href@noop {} {\bibfield  {journal} {\bibinfo  {journal} {J. Phys. G}\
  }\textbf {\bibinfo {volume} {19}},\ \bibinfo {pages} {361} (\bibinfo {year}
  {1993})},\ \Eprint {http://arxiv.org/abs/arXiv:hep-ph/9209259}
  {arXiv:hep-ph/9209259} \BibitemShut {NoStop}%
\bibitem [{\citenamefont {Bardeen}\ \emph {et~al.}(1972)\citenamefont
  {Bardeen}, \citenamefont {Gastmans},\ and\ \citenamefont
  {Lautrup}}]{Bardeen:1972vi}%
  \BibitemOpen
  \bibfield  {author} {\bibinfo {author} {\bibfnamefont {W.~A.}\ \bibnamefont
  {Bardeen}}, \bibinfo {author} {\bibfnamefont {R.}~\bibnamefont {Gastmans}}, \
  and\ \bibinfo {author} {\bibfnamefont {B.}~\bibnamefont {Lautrup}},\
  }\href@noop {} {\bibfield  {journal} {\bibinfo  {journal} {Nucl. Phys.}\
  }\textbf {\bibinfo {volume} {B46}},\ \bibinfo {pages} {319} (\bibinfo {year}
  {1972})}\BibitemShut {NoStop}%
\bibitem [{\citenamefont {Cabral-Rosetti}\ \emph {et~al.}(2000)\citenamefont
  {Cabral-Rosetti}, \citenamefont {Bernabeu}, \citenamefont {Vidal},\ and\
  \citenamefont {Zepeda}}]{CabralRosetti:1999ad}%
  \BibitemOpen
  \bibfield  {author} {\bibinfo {author} {\bibfnamefont {L.~G.}\ \bibnamefont
  {Cabral-Rosetti}}, \bibinfo {author} {\bibfnamefont {J.}~\bibnamefont
  {Bernabeu}}, \bibinfo {author} {\bibfnamefont {J.}~\bibnamefont {Vidal}}, \
  and\ \bibinfo {author} {\bibfnamefont {A.}~\bibnamefont {Zepeda}},\
  }\href@noop {} {\bibfield  {journal} {\bibinfo  {journal} {Eur. Phys. J. C}\
  }\textbf {\bibinfo {volume} {12}},\ \bibinfo {pages} {633} (\bibinfo {year}
  {2000})},\ \Eprint {http://arxiv.org/abs/arXiv:hep-ph/9907249}
  {arXiv:hep-ph/9907249} \BibitemShut {NoStop}%
\bibitem [{\citenamefont {Dvornikov}\ and\ \citenamefont
  {Studenikin}(2004{\natexlab{a}})}]{Dvornikov:2003js}%
  \BibitemOpen
  \bibfield  {author} {\bibinfo {author} {\bibfnamefont {M.}~\bibnamefont
  {Dvornikov}}\ and\ \bibinfo {author} {\bibfnamefont {A.}~\bibnamefont
  {Studenikin}},\ }\href@noop {} {\bibfield  {journal} {\bibinfo  {journal}
  {Phys. Rev. D}\ }\textbf {\bibinfo {volume} {69}},\ \bibinfo {pages} {073001}
  (\bibinfo {year} {2004}{\natexlab{a}})},\ \Eprint
  {http://arxiv.org/abs/arXiv:hep-ph/0305206} {arXiv:hep-ph/0305206}
  \BibitemShut {NoStop}%
\bibitem [{\citenamefont {Dvornikov}\ and\ \citenamefont
  {Studenikin}(2004{\natexlab{b}})}]{Dvornikov:2004sj}%
  \BibitemOpen
  \bibfield  {author} {\bibinfo {author} {\bibfnamefont {M.}~\bibnamefont
  {Dvornikov}}\ and\ \bibinfo {author} {\bibfnamefont {A.}~\bibnamefont
  {Studenikin}},\ }\href@noop {} {\bibfield  {journal} {\bibinfo  {journal} {J.
  Exp. Theor. Phys.}\ }\textbf {\bibinfo {volume} {99}},\ \bibinfo {pages}
  {254} (\bibinfo {year} {2004}{\natexlab{b}})},\ \Eprint
  {http://arxiv.org/abs/arXiv:hep-ph/0411085} {arXiv:hep-ph/0411085}
  \BibitemShut {NoStop}%
\bibitem [{\citenamefont {Davidson}\ \emph {et~al.}(2000)\citenamefont
  {Davidson}, \citenamefont {Hannestad},\ and\ \citenamefont
  {Raffelt}}]{Davidson:2000hf}%
  \BibitemOpen
  \bibfield  {author} {\bibinfo {author} {\bibfnamefont {S.}~\bibnamefont
  {Davidson}}, \bibinfo {author} {\bibfnamefont {S.}~\bibnamefont {Hannestad}},
  \ and\ \bibinfo {author} {\bibfnamefont {G.}~\bibnamefont {Raffelt}},\
  }\href@noop {} {\bibfield  {journal} {\bibinfo  {journal} {JHEP}\ }\textbf
  {\bibinfo {volume} {05}},\ \bibinfo {pages} {003} (\bibinfo {year} {2000})},\
  \Eprint {http://arxiv.org/abs/arXiv:hep-ph/0001179} {arXiv:hep-ph/0001179}
  \BibitemShut {NoStop}%
\bibitem [{\citenamefont {Fujikawa}\ and\ \citenamefont
  {Shrock}(1980)}]{Fujikawa:1980yx}%
  \BibitemOpen
  \bibfield  {author} {\bibinfo {author} {\bibfnamefont {K.}~\bibnamefont
  {Fujikawa}}\ and\ \bibinfo {author} {\bibfnamefont {R.}~\bibnamefont
  {Shrock}},\ }\href@noop {} {\bibfield  {journal} {\bibinfo  {journal} {Phys.
  Rev. Lett.}\ }\textbf {\bibinfo {volume} {45}},\ \bibinfo {pages} {963}
  (\bibinfo {year} {1980})}\BibitemShut {NoStop}%
\bibitem [{\citenamefont {Zel'dovich}(1958)}]{Zeldovich:1957zl}%
  \BibitemOpen
  \bibfield  {author} {\bibinfo {author} {\bibfnamefont {Y.}~\bibnamefont
  {Zel'dovich}},\ }\href@noop {} {\bibfield  {journal} {\bibinfo  {journal}
  {Sov. Phys. JETP}\ }\textbf {\bibinfo {volume} {6}},\ \bibinfo {pages} {1184}
  (\bibinfo {year} {1958})}\BibitemShut {NoStop}%
\bibitem [{\citenamefont {Kopeikin}\ \emph {et~al.}(1997)\citenamefont
  {Kopeikin}, \citenamefont {Mikaelyan}, \citenamefont {Sinev},\ and\
  \citenamefont {Fayans}}]{Kopeikin:1997step}%
  \BibitemOpen
  \bibfield  {author} {\bibinfo {author} {\bibfnamefont {V.~I.}\ \bibnamefont
  {Kopeikin}}, \bibinfo {author} {\bibfnamefont {L.~A.}\ \bibnamefont
  {Mikaelyan}}, \bibinfo {author} {\bibfnamefont {V.~V.}\ \bibnamefont
  {Sinev}}, \ and\ \bibinfo {author} {\bibfnamefont {S.~A.}\ \bibnamefont
  {Fayans}},\ }\href@noop {} {\bibfield  {journal} {\bibinfo  {journal} {Phys.
  At. Nucl.}\ }\textbf {\bibinfo {volume} {60}},\ \bibinfo {pages} {1859}
  (\bibinfo {year} {1997})}\BibitemShut {NoStop}%
\bibitem [{\citenamefont {Kouzakov}\ \emph {et~al.}(2011)\citenamefont
  {Kouzakov}, \citenamefont {Studenikin},\ and\ \citenamefont
  {Voloshin}}]{Kouzakov:2011vx}%
  \BibitemOpen
  \bibfield  {author} {\bibinfo {author} {\bibfnamefont {K.~A.}\ \bibnamefont
  {Kouzakov}}, \bibinfo {author} {\bibfnamefont {A.~I.}\ \bibnamefont
  {Studenikin}}, \ and\ \bibinfo {author} {\bibfnamefont {M.~B.}\ \bibnamefont
  {Voloshin}},\ }\href@noop {} {\bibfield  {journal} {\bibinfo  {journal}
  {Phys. Rev. D}\ }\textbf {\bibinfo {volume} {83}},\ \bibinfo {pages} {113001}
  (\bibinfo {year} {2011})},\ \Eprint {http://arxiv.org/abs/arXiv:1101.4878}
  {arXiv:1101.4878 [hep-ph]} \BibitemShut {NoStop}%
\bibitem [{\citenamefont {Kouzakov}\ and\ \citenamefont
  {Studenikin}(2014{\natexlab{a}})}]{Kouzakov:2014lka}%
  \BibitemOpen
  \bibfield  {author} {\bibinfo {author} {\bibfnamefont {K.~A.}\ \bibnamefont
  {Kouzakov}}\ and\ \bibinfo {author} {\bibfnamefont {A.~I.}\ \bibnamefont
  {Studenikin}},\ }\href@noop {} {\bibfield  {journal} {\bibinfo  {journal}
  {Adv. High Energy Phys.}\ }\textbf {\bibinfo {volume} {2014}},\ \bibinfo
  {pages} {569409} (\bibinfo {year} {2014}{\natexlab{a}})},\ \Eprint
  {http://arxiv.org/abs/arXiv:1406.4999} {arXiv:1406.4999 [hep-ph]}
  \BibitemShut {NoStop}%
\bibitem [{\citenamefont {Chen}\ \emph
  {et~al.}(2014{\natexlab{b}})\citenamefont {Chen} \emph
  {et~al.}}]{Chen:2013lba}%
  \BibitemOpen
  \bibfield  {author} {\bibinfo {author} {\bibfnamefont {J.-W.}\ \bibnamefont
  {Chen}} \emph {et~al.},\ }\href@noop {} {\bibfield  {journal} {\bibinfo
  {journal} {Phys. Lett. B}\ }\textbf {\bibinfo {volume} {731}},\ \bibinfo
  {pages} {159} (\bibinfo {year} {2014}{\natexlab{b}})},\ \Eprint
  {http://arxiv.org/abs/arXiv:1311.5294} {arXiv:1311.5294 [hep-ph]}
  \BibitemShut {NoStop}%
\bibitem [{\citenamefont {Kouzakov}\ and\ \citenamefont
  {Studenikin}(2014{\natexlab{b}})}]{Kouzakov:2014pepan_lett}%
  \BibitemOpen
  \bibfield  {author} {\bibinfo {author} {\bibfnamefont {K.~A.}\ \bibnamefont
  {Kouzakov}}\ and\ \bibinfo {author} {\bibfnamefont {A.~I.}\ \bibnamefont
  {Studenikin}},\ }\href@noop {} {\bibfield  {journal} {\bibinfo  {journal}
  {Phys. Part. Nucl. Lett.}\ }\textbf {\bibinfo {volume} {2014}},\ \bibinfo
  {pages} {458} (\bibinfo {year} {2014}{\natexlab{b}})},\ \Eprint
  {http://arxiv.org/abs/arXiv:1402.3786} {arXiv:1402.3786 [hep-ph]}
  \BibitemShut {NoStop}%
\bibitem [{\citenamefont {Vogel}\ and\ \citenamefont
  {Engel}(1989)}]{Vogel:1989iv}%
  \BibitemOpen
  \bibfield  {author} {\bibinfo {author} {\bibfnamefont {P.}~\bibnamefont
  {Vogel}}\ and\ \bibinfo {author} {\bibfnamefont {J.}~\bibnamefont {Engel}},\
  }\href@noop {} {\bibfield  {journal} {\bibinfo  {journal} {Phys. Rev. D}\
  }\textbf {\bibinfo {volume} {39}},\ \bibinfo {pages} {3378} (\bibinfo {year}
  {1989})}\BibitemShut {NoStop}%
\bibitem [{\citenamefont {Deniz}\ \emph {et~al.}(2010)\citenamefont {Deniz}
  \emph {et~al.}}]{Deniz:2009mu}%
  \BibitemOpen
  \bibfield  {author} {\bibinfo {author} {\bibfnamefont {M.}~\bibnamefont
  {Deniz}} \emph {et~al.} (\bibinfo {collaboration} {TEXONO Collaboration}),\
  }\href@noop {} {\bibfield  {journal} {\bibinfo  {journal} {Phys. Rev. D}\
  }\textbf {\bibinfo {volume} {81}},\ \bibinfo {pages} {072001} (\bibinfo
  {year} {2010})},\ \Eprint {http://arxiv.org/abs/arXiv:0911.1597}
  {arXiv:0911.1597 [hep-ex]} \BibitemShut {NoStop}%
\bibitem [{\citenamefont {Fabbricatore}\ \emph {et~al.}(2016)\citenamefont
  {Fabbricatore}, \citenamefont {Grigoriev},\ and\ \citenamefont
  {Studenikin}}]{Fabbricatore:2016nec}%
  \BibitemOpen
  \bibfield  {author} {\bibinfo {author} {\bibfnamefont {R.}~\bibnamefont
  {Fabbricatore}}, \bibinfo {author} {\bibfnamefont {A.}~\bibnamefont
  {Grigoriev}}, \ and\ \bibinfo {author} {\bibfnamefont {A.}~\bibnamefont
  {Studenikin}},\ }\href {\doibase 10.1088/1742-6596/718/6/062058} {\bibfield
  {journal} {\bibinfo  {journal} {J. Phys. Conf. Ser.}\ }\textbf {\bibinfo
  {volume} {718}},\ \bibinfo {pages} {062058} (\bibinfo {year} {2016})},\
  \Eprint {http://arxiv.org/abs/1604.01245} {arXiv:1604.01245 [hep-ph]}
  \BibitemShut {NoStop}%
%%CITATION = ARXIV:1604.01245;%%
\end{thebibliography}%
\end{document}